\documentclass[11pt, draftcls, onecolumn, a4paper]{IEEEtran}
\usepackage{amsmath}
\usepackage{amssymb}
\usepackage{times}
\usepackage{color}
\usepackage{graphicx}
\usepackage{cite}
\usepackage{url,subfigure}

\newcommand{\insertfig}[4]{
\begin{figure}[ht]
\centerline{\includegraphics[width=#1\columnwidth]{#2.eps}}
\caption{#3}\label{#4}\end{figure}}

\DeclareMathAlphabet{\mathsfbf}{OT1}{cmss}{sbc}{n}

\newtheorem{theorem}{Theorem}[section]

\newtheorem{corollary}{Corollary}[section]
\newtheorem{definition}{Definition}[section]

\newtheorem{remark}{Remark}[section]

\newcommand{\CC}{\mathbb{C}} 
\newcommand{\EE}{\mathbb{E}} 
\newcommand{\PP}{\mathbb{P}} 

\newcommand{\ee}{{\rm e}}
\newcommand{\jj}{{\rm j}}  
\newcommand{\ii}{{\rm i}}  
\newcommand{\dd}{{\rm\,d}} 

\newcommand{\erf}{{\rm erf}} 

\newcommand{\av}{{\bf a}}

\newcommand{\ellv}{\boldsymbol{\ell}}

\newcommand{\nv}{{\bf n}}

\newcommand{\pv}{{\bf p}}
\newcommand{\qv}{{\bf q}}

\newcommand{\sv}{{\bf s}}

\newcommand{\xv}{{\bf x}}
\newcommand{\yv}{{\bf y}}
\newcommand{\zv}{{\bf z}}

\def\Fsf{ {\sf F}}
\newcommand{\Am}{{\bf A}}

\newcommand{\Id}{{\bf I}}

\newcommand{\Vm}{{\bf V}}
\newcommand{\Xm}{{\bf X}}

\def\MSEinf{{\rm MSE}_{\infty}}

\newcommand{\Ac}{{\cal A}}
\newcommand{\Bc}{{\cal B}}
\newcommand{\Cc}{{\cal C}}

\newcommand{\Hc}{{\cal H}}

\newcommand{\Nc}{{\cal N}}

\newcommand{\Pc}{{\cal P}}

\newcommand{\Sc}{{\cal S}}
  
\newcommand{\omegav}{\boldsymbol{\omega}}

\def\tracen{\mathsf{tr}}
\def\Herm{\mathsf{^H}}
\def\Tran{\mathsf{^T}}

\def\ben{\begin{enumerate}}
\def\beq{\begin{equation}}
\def\beqa{\begin{eqnarray}}
\def\bit{\begin{itemize}}
\def\een{\end{enumerate}}
\def\eeq{\end{equation}}
\def\eeqa{\end{eqnarray}}
\def\eit{\end{itemize}}

\def\non{\nonumber\\}

\begin{document}

\title{Field Reconstruction in Sensor Networks\\
with Coverage Holes and Packet Losses}
\author{Alessandro Nordio$^\star$ and Carla-Fabiana Chiasserini$^\dag$
\vspace{3mm}\\
$^\star$ IEIIT-CNR (Italian National Research Council), Torino, Italy\\
        E-mail: alessandro.nordio@polito.it
\vspace{3mm}\\
$^\dag$ Dipartimento di Elettronica, Politecnico di Torino, Torino, Italy\\
        E-mail: chiasserini@polito.it}

\maketitle

\begin{abstract}
Environmental monitoring is often performed through a wireless sensor
network, whose nodes are randomly deployed over the geographical
region of interest. Sensors sample a physical phenomenon (the
so-called field) and send their measurements to a {\em sink} node,
which is in charge of reconstructing the field from such irregular
samples.  In this work, we focus on scenarios of practical interest
where the sensor deployment is unfeasible in certain areas of the 
geographical region,
e.g., due to terrain asperities, and the delivery of sensor
measurements to the sink may fail due to fading or to transmission collisions 
among sensors simultaneously accessing the wireless medium.
Under these conditions, we carry out an asymptotic analysis and evaluate
the quality of the estimation of a $d$-dimensional field ($d\geq 1$) 
when the sink uses linear filtering as a reconstruction technique.
Specifically, given the matrix representing the sampling system,
$\Vm$, we derive both the moments and an expression of the limiting
spectral distribution of $\Vm\Vm\Herm$, as the size of $\Vm$ goes to
infinity and its aspect ratio has a finite limit bounded away from
zero.  By using such asymptotic results, we approximate the mean
square error on the estimated field through the $\eta$-transform of
$\Vm\Vm\Herm$, and derive the sensor network performance 
under the conditions described above.
\end{abstract}

\section{Introduction}

Recently, a great deal of attention has been payed to wireless sensor
networks whose nodes sample a physical phenomenon (hereinafter
referred to as field), i.e., air temperature, light intensity, 
pollution levels or rain falls, and send their measurements to a 
central processing unit
(or {\em sink} node).  The sink is in charge of reconstructing the
sensed field: if the field can be approximated as bandlimited in the
time and space domain, then an estimate of the discrete spectrum can
be obtained.

However, the sensors measurements typically represent an irregular
sampling of the field of interest, thus the sink operates based on a
set of field samples that are not regularly spaced in the time and
space domain.  The reasons for such an irregular sampling are multifold.  
(i) The sensors may be
irregularly deployed in the geographical region of interest, either due
to the adopted deployment procedure (e.g., sensors thrown out of an
airplane \cite{smartdust}), or due to the presence of terrain
asperities and obstacles.  (ii) The transmission of the measurements
from the sensors to the central controller may fail due to bad channel
propagation conditions (e.g., fading), or because collisions occur among the
transmissions by sensors simultaneously attempting to access the channel. 
In this case, although the sample has been collected
by the sensor, it will not be delivered to the central controller.
(iii) The sensors may enter a low-power operational state (sleep
mode), in order to save energy~\cite{Perillo04,Willett04}.  While in
sleep mode, the nodes neither perform sensing operations nor
transmit/receive any measurement.  (iv) The sensors may be loosely
synchronized, hence sense the field at different time instants.

Clearly, sampling irregularities may result in a degradation of the
reconstructed signal~\cite{Marvasti}.  The work in~\cite{Ganesan03}
investigates this issue in the context of sensor networks. Other
interesting studies can be found in~\cite{Zhao} and~\cite{Early}, just
to name a few, which address the perturbations of regular sampling in
shift-invariant spaces~\cite{Zhao} and the reconstruction of
irregularly sampled images in presence of measure noise~\cite{Early}.

In this work, our objective is to evaluate the performance of the
field reconstruction when the coordinates in the $d$-dimensional
domain of the field samples, which reach the sink node, are randomly,
independently distributed and the sensors measurements are noisy.  We
take as performance metric the mean square error (MSE) on the
reconstructed field. As a reconstruction technique, we use linear
filtering and we adopt the filter that minimizes the MSE (i.e., the
LMMSE filter)~\cite{TSP1,TSP2,TSP3}.
The matrix representing the sampling system, in the following denoted
by $\Vm$, results to be a $d$-fold Vandermonde matrix\footnote{An $n
  \times m$ matrix $\Xm$ is Vandermonde if its $(i,j)-$th entry,
  $(\Xm)_{ij}$ can be written as $(\Xm)_{ij}=x_j^i$, $i=0,\ldots,n-1,
  \, j=1,\ldots,m$.}.  By drawing on the results
in~\cite{TSP2,RyanDebbah}, we derive both the moments and an
expression of the limiting spectral distribution (LSD) of
$\Vm\Vm\Herm$, as the size of $\Vm$ goes to infinity and its aspect
ratio has a finite limit bounded away from zero.  Then, by using such
an asymptotic model, we approximate the MSE on the reconstructed field
through the $\eta$-transform~\cite{tutorial} of $\Vm\Vm\Herm$, and
derive an expression for it.
We apply our results to the study of network scenarios of practical
interest, such as sensor sensor deployments with coverage holes,
communication in presence of a fading channel, massively dense networks~\cite{massdense1, massdense2}, 
and networks using contention-based channel access
techniques~\cite{802.15.4}.

The rest of the paper is organized as follows.
Section~\ref{sec:related} reviews previous work, while  
Section~\ref{sec:system} describes the system model
under study.
In Section~\ref{sec:preliminaries}, we first provide some useful definitions
and introduce our performance
metric, then  we recall previous results on which we build our
analysis.  
In Section~\ref{sec:results-vandermonde}, we derive asymptotic results
concerning the moments and the LSD of $\Vm\Vm\Herm$. Such results are applied to
different practical scenarios in Section~\ref{sec:applications}.
Finally, Section~\ref{sec:conclusions} concludes the paper.

\section{Related Work\label{sec:related}}
In the context of sensor networks, several works \cite{CristescuVetterli,Poor,Vuran04,Rachlin1} have studied the field reconstruction at the
sink node in presence of spatial and temporal correlation among 
sensor measurements.  In
particular, in~\cite{Rachlin1} the observed field is a discrete vector
of target positions and sensor observations are dependent. By modeling
the sensor network as a channel encoder and exploiting some concepts
from coding theory, the network capacity, defined as the maximum value of the
ratio of the target positions to the number of sensors, 
is studied as a function of
the noise, the sensing function and the sensor connectivity level.
 
The paper by Dong and Tong~\cite{DongTong} considers a dense sensor
network where a MAC protocol is responsible to collect samples from
network nodes.  The work analyzes the impact of deterministic and
random data collection strategies on the quality of field
reconstruction.  As a performance measure, the maximum of the
reconstruction square error over the sensed field is employed, as
opposed to our work where the mean square error is considered. Also,
in \cite{DongTong} the field is a Gaussian random process and 
the sink always receives a sufficiently large number of samples
so as to reconstruct the field with the required accuracy.

The problem of reconstructing a bandlimited field from a 
set of irregular samples at unknown locations, instead, has been addressed
in~\cite{Marziliano00}. There, the field is oversampled by
irregularly spaced sensors; sensor positions are unknown but always
equal to an integer multiple of the sampling interval. Different
solution methods are proposed, and the conditions for which there
exist multiple solutions or a unique solution are discussed.
Differently from~\cite{Marziliano00}, we assume that the sink can
either acquire or estimate the sensor locations and that the
coordinates of the sampling points are randomly located over a finite
$d$-dimensional domain.

As for previous results on Vandermonde matrices, in~\cite{RyanDebbah}
Ryan and Debbah considered a Vandermonde matrix $\Vm$ with $d=1$ and
complex exponential entries, whose phases are i.i.d. with continuous
distribution. Under such hypothesis, they obtained the important
results that, given the phases distribution, the moments of
$\Vm\Vm\Herm$ can be derived once the moments for the case with
uniformly distributed phases are known.  Also, a method for computing
the moments of sums and products of Vandermonde matrices, for the
non-folded case (i.e., $d=1$), has recently appeared
in~\cite{ryan-2009}; further insights on the extremal eigenvalues
behavior, still for the case of non-folded Vandermonde matrices, can
be found in~\cite{Tucci2010}. Moreover, in~\cite{TSP2} it has been
shown that the LSD of $\Vm\Vm\Herm$ converges to the
Mar\v{c}enko-Pastur distribution~\cite{MarcenkoPastur} when $\Vm$ is
$d$-fold Vandermonde with uniformly distributed phases and
$d\rightarrow\infty$.

Note that, with respect to previous studies on Vandermonde matrices with entries
that are randomly distributed on the complex unit circle, 
in this work we obtain results on
the LSD of $\Vm\Vm\Herm$ where the entries of $\Vm$ have phases drawn
from a {\em generic continuous distribution}. By relying on the
results in~\cite{TSP2,RyanDebbah}, we show that such an LSD can be related
to that of $\Vm\Vm\Herm$ when the phases of $\Vm$ are {\em uniformly}
distributed on the complex unit circle.  We also provide some
numerical results that show the validity of our analysis.  
To our knowledge, these results have not been previously derived.
We then apply them to the study of several practical scenarios in the 
context of sensor networks, 
although our findings can be useful for the study of other aspects 
of communications as well~\cite{RyanDebbah}.

\section{Network Model\label{sec:system}}

We consider a network composed of $m$ wireless sensors, which measure
the value of a spatially-finite physical field defined over $d$
dimensions, ($d \ge 1$).  We denote by
$\Hc=\left[-\frac{1}{2},\frac{1}{2}\right)^d$ the hypercube over which
  the sampling points fall, and we assume that the sampling points are
  i.i.d. randomly distributed variables, whose value is known to the
  sink node.  Note that this is a fair assumption, as one can think of
  sensor nodes randomly deployed over the geographical region that has
  to be monitored, or, even in the case where the network topology is
  intended to have a regular structure, the actual node deployment may
  turn out to be random due to obstacles or terrain asperities.  In
  addition, now and then the sensors may enter a low-operational mode
  (hence become inactive) in order to save energy, and they may be
  loosely synchronized.  All the above conditions yield a set of
  randomly distributed samples of the field under observation, in both
  the time and the space domain~\cite{Ganesan03}.

By truncating its Fourier series expansion, a physical field defined
over $d$ dimensions and with finite energy can be approximated in the
region $\Hc$ as \cite{TSP2}
  \begin{equation}
  \label{eq:blsig}
n^{-d/2}\sum_{\ellv}
  a_{\nu(\ellv)} \ee^{\jj 2\pi \ellv\Tran \xv }
   \end{equation}
where $n$ is the approximate one-sided bandwidth (per dimension) of
the field, $\ellv=[\ell_1,\ldots,\ell_d]\Tran$ is a vector of
integers, with $\ell_k=0,\ldots,n-1$, $k=1,\ldots,d$. The coefficient
$n^{-d/2}$ is a normalization factor and the function
\[\nu(\ellv) = \sum_{j=1}^d n^{j-1} \ell_j,\]
maps uniquely the vector $\ellv$ over $[0,n^d-1]$.  
$a_{\nu(\ellv)}$ denotes the  $\nu(\ellv)$-th entry of the 
vector $\av$ of size $n^d$, which represents the approximated field spectrum, 
while the real vectors $\xv_q$, $q=1,\ldots,m$
represent the coordinates of the $d$-dimensional sampling points. In
this work, we assume that $\xv_q$, $q=1,\ldots,m$, are i.i.d. random
vectors having a generic continuous distribution $f_{x}(\zv)$,
$\zv\in \Hc$.  In the specific case where $\xv_q$ are i.i.d with i.i.d.
entries $x_{qj}, j=1,\ldots,d$, uniformly distributed in $[-1/2,1/2)$,
  we denote the distribution of $\xv_q$ by $f_u(\zv)$.

The coordinates of the $d$-dimensional sampling points, however, 
are known to the sink node, because
{\em (i)} either sensors are located at pre-defined positions or their
position can be estimated through a localization
technique~\cite{Moore04}, and {\em (ii)} the sampling time is either
periodic or included in the information sent to the sink.

Now, let $\sv=[s(\xv_1),\ldots,s(\xv_m)]^{\rm T}$ be the values of
the samples at $[\xv_1,\ldots,\xv_m]$, respectively.
Following~\cite{TSP1,TSP2}, we can write the vector $\sv$ as a
function of the field spectrum:
\begin{equation}
\label{eq:s}
\sv= \beta_{n,m}^{-1/2}\Vm\Herm\av 
\end{equation}
where $\Vm$ is the $n^d \times m$ $d$-fold Vandermonde matrix with
entries
\begin{equation}
\label{def:V_multifold}
\left(\Vm_{\nu(\ellv),q}\right)=m^{-1/2}\exp\left(-2\pi \ii \ellv\Tran
  \xv_q\right)
\end{equation}  
randomly distributed on the complex circle of radius $m^{-1/2}$, and
$\beta_{n,m}$ is the ratio of the rows to the columns of $\Vm$, i.e., 
\[ \beta_{n,m} = \frac{n^d}{m} \,.\]

In general, the entries of $\av$ can be correlated with covariance
matrix $\EE[\av \av\Herm]$. However, in the following, we
restrict our attention to the class of fields characterized by
$\EE[\av \av\Herm]= \sigma^2_a\Id$.

In the case where the sensor measurements, $\pv = [p_1,\ldots, p_m]\Tran$, are
noisy, then the relation between the sensor samples and the 
approximated field spectrum can be written as:
\begin{equation}
\pv = \sv + \nv = \beta_{n,m}^{-1/2}\Vm\Herm\av + \nv
\label{eq:p}
\end{equation}
where $\nv$ is a $m$-size, zero-mean random vector  representing the noise.
Here, we assume a white noise, i.e., with covariance matrix 
$\EE[\nv\nv\Herm] = \sigma^2_n\Id_{m}$.  
Note that the additive white noise affecting the
sensor measurements may be due to quantization, round-off errors or
quality of the sensing device.

\section{Preliminaries\label{sec:preliminaries}} 

In this section, we report some definitions and previous results that 
are useful for our study.

\subsection{Useful definitions} 
Let us consider an $n \times n$ non-negative definite random matrix $\Am$,
whose eigenvalues are denoted by $\lambda_{\Am,1}, \ldots, \lambda_{\Am, n}$.

\begin{definition}
The average empirical cumulative distribution of the eigenvalues of
$\Am$ is defined as $\Fsf_{\lambda_\Am}^{(n)} (z)= \frac{1}{n}
\sum_{i=1}^n \EE \left [ 1\{\lambda_{\Am,i} \leq z \} \right]$, where
the superscript $(n)$ indicates that we refer to a system with size
$n$ and $1 \{\cdot\}$ is the indicator function.  If
$\Fsf_{\lambda_\Am}^{(n)}(z)$ converges as $n \rightarrow \infty$,
then $\lim_{n \rightarrow \infty}\Fsf_{\lambda_\Am}^{(n)}(z) =
\Fsf_{\lambda_\Am}(z)$.  The corresponding limiting probability
density function, or limiting spectral distribution (LSD), is denoted
by $f_{\lambda_\Am}(\cdot)$.
\end{definition}

\begin{definition}
The $\eta$-transform of $\Am$ is given by:
\begin{equation}
\label{eq:eta-T}
\eta^{(n)}_\Am(\gamma)
=  \EE\left[\tracen\left\{\left(\gamma \Am+\Id\right)^{-1} \right\}\right]  = \EE\left[\frac{1}{n}\sum_{i=1}^n\frac{1}{\gamma\lambda_{\Am,i} +1}\right] 
\end{equation}
where $\tracen\{\cdot\}$ is the normalized matrix trace operator 
and $\gamma$ is a non-negative real number.  If
$\eta_{\Am}^{(n)}(\gamma)$ converges as $n \rightarrow \infty$, then
the corresponding limit is $\eta_{\Am}(\gamma) = \EE[(\gamma
  \lambda_\Am+1)^{-1}]$~\cite[p. 40]{tutorial},  where 
$\lambda_\Am$ is the generic asymptotic eigenvalue of $\Am$, 
whose distribution is $f_{\lambda_\Am}(z)$, and the average is
computed with respect to $\lambda_\Am$~\cite{tutorial}.
\end{definition}

Next, consider the matrix $\Vm$ as defined in (\ref{def:V_multifold}) 
and that  the LMMSE filter is used for field reconstruction.
Then, the estimate of the unknown vector
$\av$ in~(\ref{eq:p}), given $\yv$ and $\Vm$, is obtained by computing
$\hat{\av} = \EE[\av\pv\Herm] \EE[\pv\pv\Herm]^{-1} \pv$. Through easy
computations and using the Sherman-Morrison-Woodbury identity, 
we can obtain the MSE as 
\begin{equation}
{\rm MSE}^{(n)}
=\sigma_a^{-2}\EE\left[\tracen\left\{\left(\sigma_n^{-2}\beta_{n,m}^{-1}\Vm\Vm\Herm+\sigma_a^{-2}\Id\right)^{-1}\right\}\right] = \eta^{(n)}_{\Vm\Vm\Herm}\left(\frac{\gamma}{\beta_{n,m}}\right)
\label{eq:MSE_n}
\end{equation}
where  $\gamma=\sigma^2_a/\sigma^2_n$ denotes the signal-to-noise
ratio on the sensor measurements, and we employed the definition of the
$\eta$-transform given in~(\ref{eq:eta-T}).

Next, we approximate the MSE of the finite size system in~(\ref{eq:p})
through an asymptotic model, which assumes the size of $\Vm$ to grow
to infinity while the ratio of its number of rows to its number of
columns tends to a finite limit, $\beta$, greater than zero, i.e., we
assume
\[ \lim_{n,m\rightarrow \infty} \beta_{n,m} = \beta\] 

Indeed, in our recent works~\cite{TSP1,TSP2,TSP3} it was shown that this
asymptotic model provides a tight approximation of the MSE of the
finite size system, already for small values of $n$ and $m$.
Under these conditions, we therefore define the asymptotic expression
of the MSE as~\cite{TSP3}:
\begin{equation}
\MSEinf = \lim_{n,m\rightarrow \infty} {\rm MSE}^{(n)} =
\eta_{\Vm\Vm\Herm}\left(\gamma/\beta\right)
\label{eq:mseinf}
\end{equation}
if the limit exists.

\subsection{Previous results}

Vandermonde matrices have been studied in a number of recent
works~\cite{TSP1,TSP2,TSP3,RyanDebbah}. Specifically,~\cite{TSP2}
considered the case where the vectors $\xv_q$ are i.i.d., for
$q=1,\ldots,m$, and their entries, $x_{qj}$ are i.i.d. random
variables with uniform distribution in $[-1/2,1/2)$. The work there
studied the eigenvalue distribution 
  of $\Vm\Vm\Herm$ for both finite and infinite (i.e., $m,n\rightarrow
  \infty$) matrix size.  Although an explicit expression of such LSD
  is still unknown,~\cite{TSP2} provided an algorithm to compute its
  moments of any order in closed form.

Indeed, as $n,m\rightarrow \infty$ with $\beta_{n,m}=n^d/m$ having a
finite limit $\beta>0$, in~\cite{TSP2} it was
shown that the $p$-th moment of the generic asymptotic eigenvalue of 
$\Vm\Vm\Herm$, denoted by $\lambda$, is given by
\[ M_{p,d,\beta,u} = \int z^p f_{\lambda,u}(d,\beta,z) \dd z = 
\sum_{k=1}^p \beta^{p-k} \sum_{\omegav \in
  \Omega_{p,k}}v(\omegav)^d \] where $f_{\lambda,u}(d,\beta,z)$ 
represents the distribution of  $\lambda$.  Moreover, $\Omega_{p,k}$
is the set of partitions of the set $\Pc=\{1,2,\ldots,p\}$ in $k$
subsets, and $v(\omegav) \in (0,1]$, $\omegav \in \Omega_{p,k}$ is a
rational number that can be analytically computed from $\omegav$
following the procedure described in~\cite{TSP2}. The
subscript $u$ in $M_{p,d,\beta,u}$ and $f_{\lambda,u}(d,\beta,z)$
indicates that a uniform distribution of the entries of $\xv_q$ is
considered in the matrix $\Vm$.

In~\cite{TSP2} it was also shown that when $n,m,d\rightarrow \infty$,
with $\beta_{n,m}=n^d/m$ having a finite limit $\beta>0$, the eigenvalue
distribution $f_{\lambda,u}(d,\beta,z)$ converges to the
Mar\v{c}cenko-Pastur law~\cite{MarcenkoPastur}. A similar
result~\cite{TSP3} also applies when the vectors $\xv_q$
($q=1,\ldots,m$) are independent but not i.i.d., with equally spaced
averages.

More recently, Ryan and Debbah in~\cite{RyanDebbah} considered $d=1$
and the case where the random variables $x_{q1}$, $q=1,\ldots,m$, are
i.i.d. with continuous distribution $f_x(z)$, $0 \le z < 1$. Under
such hypothesis, it was shown that the asymptotic moments of
$\Vm\Vm\Herm$ can be written as
 \begin{eqnarray}
M_{p,1,\beta,x} = \sum_{k=1}^p I_k \beta^{p-k} \sum_{\omegav \in \Omega_{p,k}}v(\omegav)
\label{eq:moments_1D}
\end{eqnarray}
where the terms $I_k$ depend on the phase distribution $f_x(z)$ and
are given by
\[ I_k = \int_0^1 f_x(z)^k \dd z \]
for $k\ge 1$.  The subscript $x$ in $M_{p,1,\beta,x}$ indicates
that in the matrix $\Vm$ the random variables $x_{q1}$ have
a generic continuous distribution $f_x(z)$.  
Note that for the uniform distribution we
have $I_k=1$, for all $k$. The important result
in~(\ref{eq:moments_1D}) states that, given $\beta$, if the moments of
$\Vm\Vm\Herm$ are known for uniformly distributed phases, they can be
readily obtained for any continuous phase distribution $f_x(z)$.

\section{Vandermonde matrices with generic phase distribution}
\label{sec:results-vandermonde} 

In this work, we extend the above results by considering a sampling
system defined over $d\ge 1$ dimensions with nonuniform sample
distribution, where samples may be irregularly spaced in the time and
spatial domains, as it occurs in wireless sensor networks.  Being our
goal the estimation of the quality of the reconstructed field, we aim
at deriving the asymptotic MSE (i.e.,
$\eta_{\Vm\Vm\Herm}(\gamma/\beta)$).

We start by considering a generic continuous distribution,
$f_x(\zv)$, $\zv\in \Hc$ of the samples measured by the sensors over
the $d$-dimensional domain.  We state the theorem below, which gives
the asymptotic expression of the generic moment of $\Vm\Vm\Herm$, for
$d\geq1$.

\begin{theorem}
\label{th:1}
Let $\Vm$ a $d$-fold $n^d\times m$ Vandermonde matrix with entries
given by~(\ref{def:V_multifold}) where the vectors $\xv_q$,
$q=1,\ldots,m$, are i.i.d. and have continuous distribution
$f_x(\zv)$.  Then, for $n,m\rightarrow \infty$, with $\beta_{n,m}=n^d/m$ 
having a finite limit $\beta>0$, 
the $p$-th moment of $\Vm\Vm\Herm$ is given by
\begin{eqnarray}
M_{p,d,\beta,x} = \sum_{k=1}^p \beta^{p-k} I_k \sum_{\omegav \in
  \Omega_{p,k}}v(\omegav)^d
\end{eqnarray}
where $I_k = \int_\Hc f_x(\zv)^k \dd \zv$ and the terms $v(\omegav)$
are defined as in~\cite{TSP2}. 

The proof is given in Appendix~\ref{app:th1}.
\end{theorem}

Using Theorem~\ref{th:1} and the definition of $I_k$, it it possible
to show the theorem below, which provides the LSD of $\Vm\Vm\Herm$.
\begin{theorem}
\label{th:2}
Let
\begin{itemize}
\item $\Vm$ be a $d$-fold $n^d\times m$ Vandermonde matrix with
  entries given by~(\ref{def:V_multifold}) where the vectors $\xv_q$,
  $q=1,\ldots,m$, are i.i.d. and have continuous distribution
  $f_x(\zv)$, $\zv \in \Hc$
\item $\Ac$ be the set where $f_x(\zv)$ is strictly positive,
  i.e., $\Ac = \left\{ \zv \in \Hc | f_x(\zv)>0 \right\}$ 
\item the cumulative density function
\begin{equation}
G_x(y) = \frac{1}{|\Ac|}\left|\left\{\zv \in \Ac \left|  f_x(\zv) \le y\right. \right\}\right|
\label{eq:G}
\end{equation}
defined%
\footnote{$|\Sc|$ denotes the measure of the set $\Sc$} for $y >0$ and
let $g_x(y)$ be its corresponding probability density function.
\end{itemize}
Then, the LSD of $\Vm\Vm\Herm$, for $n,m\rightarrow \infty$ with
$\beta_{n,m}=n^d/m$ having a finite limit $\beta>0$, is given by
\begin{equation}
f_{\lambda,x}(d,\beta,z) = \left(1-|\Ac|\right)\delta(z) + |\Ac|\int_0^{\infty}\frac{g_x(y)}{y}f_{\lambda,u}\left(d,\frac{\beta}{y},\frac{z}{y}\right)\dd y
\label{eq:th2}
\end{equation}
\end{theorem}
\begin{IEEEproof}
The proof can be found in Appendix~\ref{app:th2}.
\end{IEEEproof}

From Theorem~\ref{th:2}, the corollary below follows.
\begin{corollary}
\label{cor-scal}
Consider $f_{x}(\zv)$ such that  $f_{x}(\zv)>0$ $\forall\zv \in \Hc$.
Then, let us denote
by $f_{x'}(\zv)$ a scaled version of this function, so that 
\begin{equation}
f_{x'}(\zv) = \left \{ \begin{array}{ll}
\frac{1}{|\Cc|}f_x\left(\frac{\zv}{|\Cc |}\right) & \zv \in \Cc \\
0 & \zv \in \Hc\setminus\Cc
\end{array} \right.
\label{eq:fxi'_scaled}
\end{equation}
where $\Cc \subset \Hc$.  It can be shown that
\begin{equation} f_{\lambda,x'}(d,\beta,z) = 
\left(1-c\right)\delta(z)+c^2f_{\lambda,x}(d,c \beta, cz) 
\label{eq:f-scal}
\end{equation} 
where $c=|C|$.
\end{corollary}
\begin{IEEEproof}
The proof can be found in Appendix~\ref{app:cor-scal}.
\end{IEEEproof}

As an example of the result given in Corollary~\ref{cor-scal}, consider
that a unidimensional ($d=1$) sensor network monitors the segment
$\Hc=[-1/2,1/2]$. Due to terrain irregularities and obstacles, nodes
are deployed with uniform distribution only in the range $[-c/2,c/2]$
(with $c\in [0,1)$).  We therefore have $f_{x'}(z)= 1/c$ for
  $-c/2\le z \le c/2$ and 0 elsewhere.  Moreover,
  $f_{\lambda,x}(1,\beta,z) = f_{\lambda,u}(1,\beta,z)$. The
  expression of $f_{\lambda,x'}(1,\beta,z)$ is given
  by~(\ref{eq:f-scal}), by replacing $d=1$ and the
  subscript $x$ with the subscript $u$. 

\begin{figure}
\centering 
\subfigure[]{\label{fig:beta08_x08} 
\includegraphics[width=0.45\columnwidth]{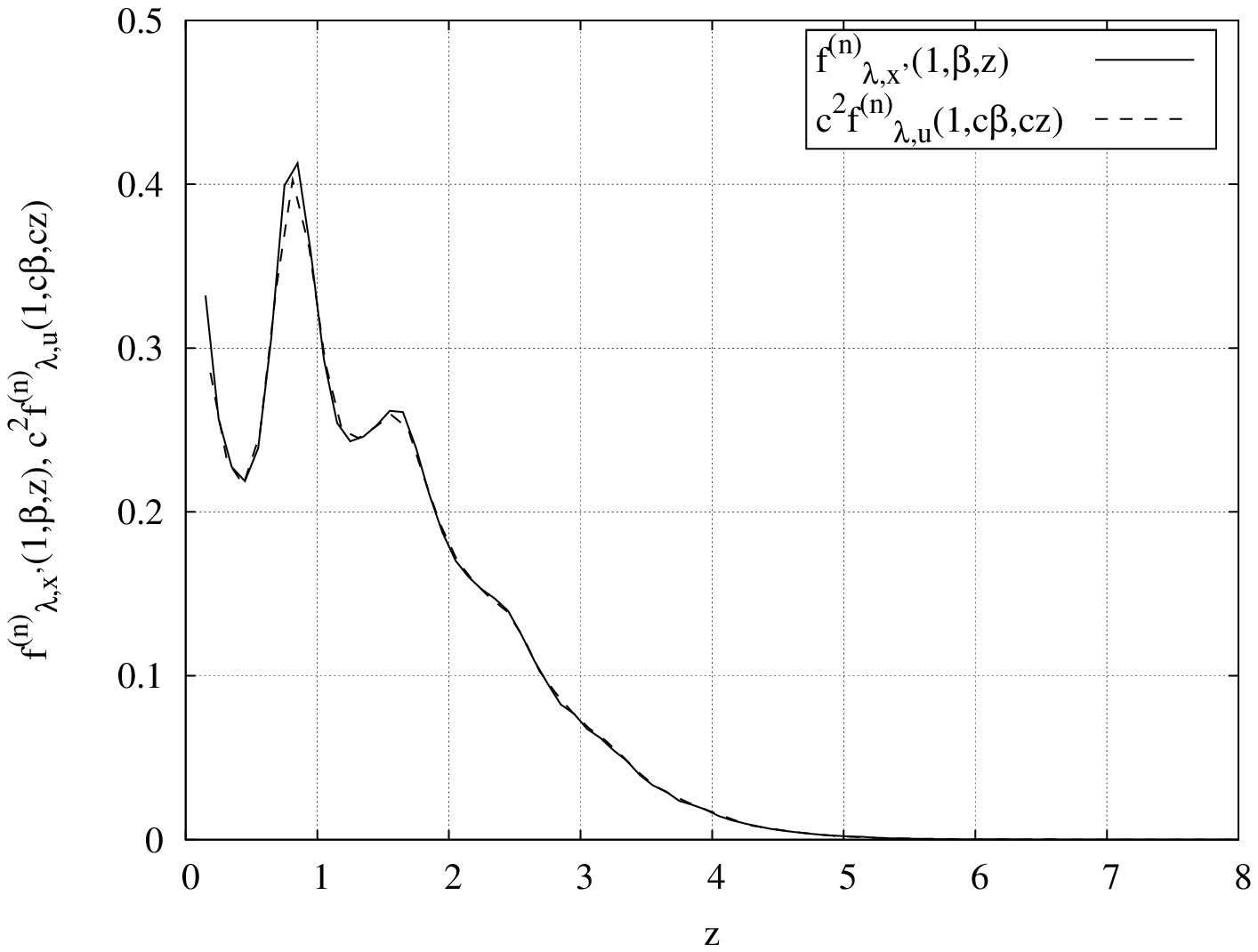}
}
\hspace{0.05\columnwidth}
\subfigure[]{\label{fig:beta02_x05}
\includegraphics[width=0.45\columnwidth]{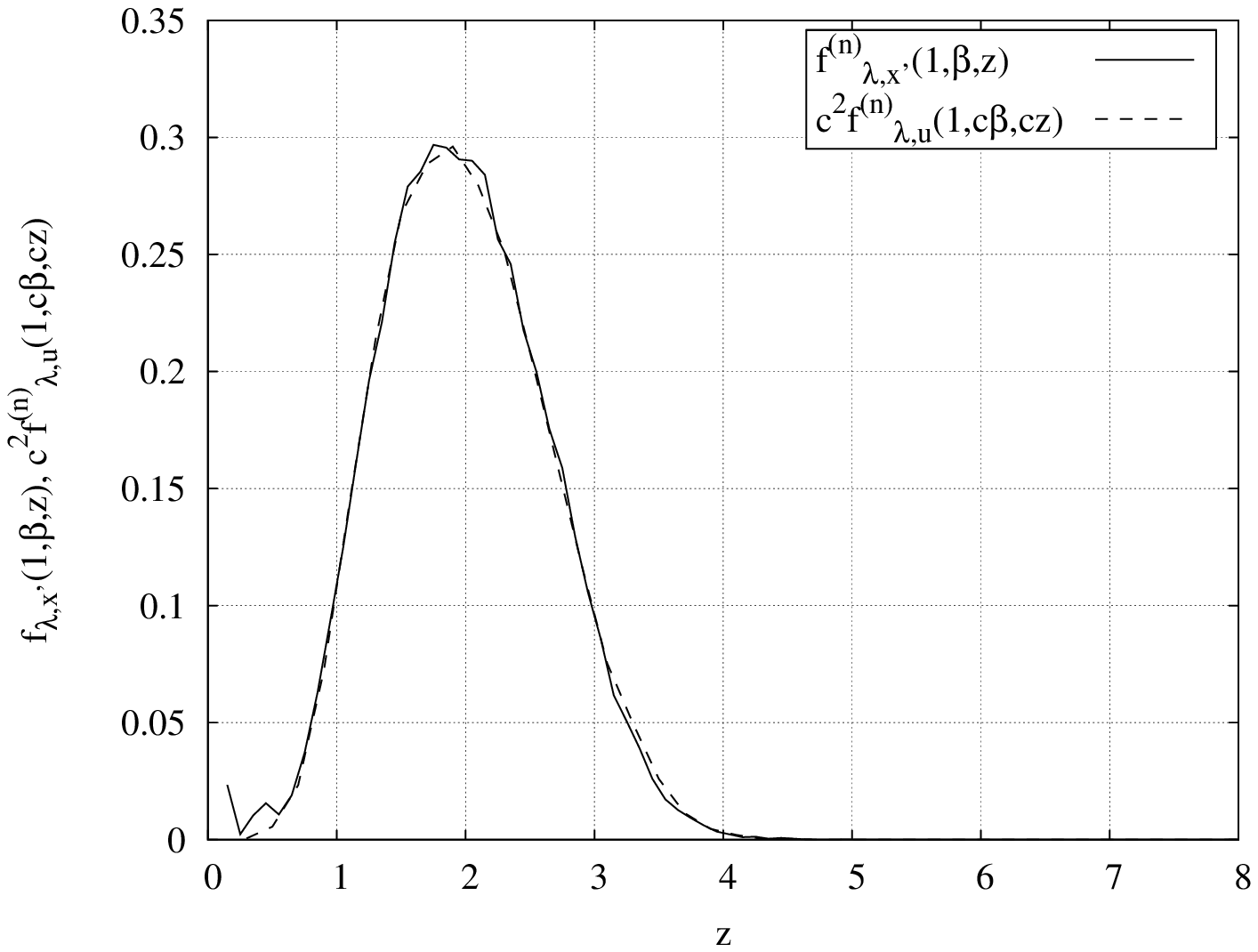}
}
\caption{Comparison between the curve representing 
$f^{(n)}_{\lambda,x'}(1,\beta_{n,m},z)$ and the one representing
the empirical function $c^2f^{(n)}_{\lambda,u}(d,\beta_{n,m} c,zc)$.
(a) $\beta_{n,m}=0.8$, $c=0.8$ and (b) $\beta_{n,m}=0.2$, $c=0.5$.}
\end{figure}

This result is well
  supported by simulations as shown in Figures~\ref{fig:beta08_x08}
  and~\ref{fig:beta02_x05}.
In the plots, we compare the asymptotic empirical spectral distribution 
(AESD) $f^{(n)}_{\lambda,x'}$ and
$f^{(n)}_{\lambda,u}$ instead of the LSDs $f_{\lambda,x'}$ and
$f_{\lambda,u}$ since an analytic expression of $f_{\lambda,u}$ is
still unknown.  However, in \cite{TSP1,TSP2,TSP3} it is shown
that, already for small values of $n$, the AESD 
$f^{(n)}_{\lambda,u}$ appears to rapidly
converge to a limiting distribution.
Figure~\ref{fig:beta08_x08} refers to the case $\beta_{n,m}=0.8$
and $c=0.8$. The solid and dashed lines represent, respectively, the
functions $f^{(n)}_{\lambda,x'}(1,\beta,z)$ and
$c^2f^{(n)}_{\lambda,u}(1,\beta_{n,m} c,zc)$, for $n=100$.  Note that
the probability mass of $f^{(n)}_{\lambda,x'}(1,\beta_{n,m},z)$ at $z=0$ is not
shown for simplicity. Similarly, Figure~\ref{fig:beta02_x05} shows
the case $\beta_{n,m}=0.2$ and $c=0.5$. As evident from these plots,
the match between the two functions is excellent for any parameter
setting, thus supporting our findings.

Since we are  interested in evaluating the MSE, taking into account 
the result in (\ref{eq:mseinf}), we now apply the definition of the 
$\eta$-transform to~(\ref{eq:th2}).
The corollary below immediately follows.
\begin{corollary}
\label{cor2}
The $\eta$-transform of $\Vm\Vm\Herm$ is given by
 \begin{equation} 
\eta_{x}(d,\beta,\gamma) = 1-|\Ac|+|\Ac|\int_0^{\infty}g_x(y) \eta_{u}\left(d,\frac{\beta}{y},\gamma y\right)\dd y 
\label{eq:eta}
\end{equation}
hence, the asymptotic MSE on the reconstructed field, 
defined in~(\ref{eq:mseinf}), is given by 
 \begin{equation} 
\MSEinf=\eta_{x}(d,\beta,\frac{\gamma}{\beta}) = 1-|\Ac|+|\Ac|
\int_0^{\infty}g_x(y) \eta_{u}\left(d,\frac{\beta}{y},\frac{\gamma}{\beta} y\right)\dd y 
\label{eq:MSEinf-eta}
\end{equation}
\end{corollary}
\begin{IEEEproof}
The proof can be found in Appendix~\ref{app:cor2}.
\end{IEEEproof}
In~(\ref{eq:eta}), in order to avoid a heavy notation we referred to
$\eta_{\Vm\Vm\Herm}(d,\beta,\gamma)$ as $\eta_{x}(d,\beta,\gamma)$
when the phases of the entries of $\Vm$ follow a generic random
continuous distribution, while $\eta_{u}(d,\beta,\gamma)$ refers to
the case where the phases are
uniformly distributed.

\begin{remark}
Since $g_x(y)>0$ and $\eta_u(d,\beta/y,\gamma y/\beta)>0$, the integral in
the right hand side of~(\ref{eq:eta}) is positive, then
$\eta_x(d,\beta,\gamma/\beta)> 1-|\Ac|$.  It follows that the MSE is
lower-bounded by the measure of the total area where the probability
of finding a sensor is zero. This clearly suggests
that, in order to obtain a
good quality of the field reconstructed at the sink node, 
this area must be a small fraction of the region under observation.
\end{remark}

Next, we observe that, in the case of massively dense networks
where the number of sampling sensors is much larger than the number of
harmonics considered in the approximated field, i.e., $\beta \ll 1$,
an interesting result holds:
\begin{corollary}
\label{cor1}
Let $\Ac$ be the set where $f_{x}(\zv)$ is strictly positive; then
\begin{equation}
 \lim_{\beta\rightarrow 0} f_{\lambda,x}(d,\beta,z) = \left(1-|\Ac|\right)\delta(z)+ |\Ac| g_x(z)
 \label{eq:cor1}
\end{equation}
\end{corollary}
\begin{IEEEproof}
The proof can be found in Appendix~\ref{app:cor1}.
\end{IEEEproof}
Thus, as evident from 
Corollary~\ref{cor1}, for the limit of $\beta
\rightarrow 0$, the LSD of $\Vm\Vm\Herm$ is the density of the density
of the phase distribution $f_x(\zv)$.

Furthermore, for massively dense networks, we have:
\begin{corollary}
\label{cor3}
Let $\Ac$ be the set where $f_{x}(\zv)$ is strictly positive; then 
\begin{eqnarray} 
\lim_{\beta\rightarrow 0} \eta_x(d,\beta,\gamma/\beta) &=& 1-|\Ac| \label{eq:cor3.2}
\end{eqnarray}
\end{corollary}
\begin{IEEEproof}
The proof can be found in Appendix~\ref{app:cor3}.
\end{IEEEproof}
\begin{remark}
The result in~(\ref{eq:cor3.2}) shows that even for massively dense
networks $1-|\Ac|$ is the minimum achievable $\MSEinf$, when an area
$\Ac$ cannot covered by sensors.
\end{remark}


\section{Field reconstruction in presence of losses\label{sec:applications}}

Here, we provide examples of how our results can be used in 
wireless sensor networks to 
investigate the impact of a random distribution of the coordinates
of the sampling points on the quality of the reconstructed field. 
In particular, we first consider a wireless channel 
affected by fading, and then  the effects of contention-based
channel access.

\subsection{Sensor network performance with fading communication channel}
We consider a wireless sensor network whose nodes are uniformly
distributed over a geographical region.  Without loss of generality,
we assume a square region of unitary side ($d=2$,
$\Hc=[-1/2,+1/2]^2$), where the sink is located at the center and has
coordinates $(z_1,z_2)=(0,0)$.  Through direct transmissions, the
sensors periodically send messages to the sink, including their
measurements.  At every sample period, a sensor message is correctly
received at the sink if its signal-to-noise ratio (SNR) exceeds a
threshold $\tau$.  The communication channel is assumed to be affected
by slow fading and to be stationary over the message duration.

Let $d$ be the distance between a generic sensor and the sink. Then, the signal
to noise ratio at the receiver is given by 
\[ {\rm SNR}(d) = s |h|^2d^{-2} \]
where $h\sim \Nc_\CC(0,1)$ is a circularly symmetric
Gaussian complex random variable representing the channel gain, and
 $s$ is the signal to noise ratio in the absence of fading and
when the sensor-sink  distance is $d=1$.

The probability that a message is correctly received at the sink is
given by
\begin{eqnarray} 
 \PP({\rm SNR}(d) > \tau) 
&=& \PP\left(|h|^2 > \tau\frac{d^2}{s}\right) \non
&=& 1-F_{|h|^2}\left( \tau\frac{d^2}{s} \right) \non
&=& \exp\left(-a d^2\right)
\label{eq:SNR}
\end{eqnarray}
with $a=\tau/s$ and $F_{|h|^2}(z) = 1-\ee^{-z}$ being the cumulative density
function of $|h|^2$.

The probability density $f_{x}(z_1,z_2)$ corresponding to
sensors at distance $d=\sqrt{z_1^2+z_2^2}$, $-1/2\le
z_1,z_2\le 1/2$ from the sink and successfully sending a message 
is then given by
\[ f_{x}(z_1,z_2) = \frac{f_u(z_1,z_2) \PP\left({\rm SNR}\left(\sqrt{z_1^2+z_2^2}\right) > \tau\right)}{\iint_{\Hc} f_u(z_1,z_2) \PP\left({\rm SNR}\left(\sqrt{z_1^2+z_2^2}\right) > \tau\right)\dd z_1 \dd z_2} \]
where $f_u(z_1,z_2)=1\, \forall z_1,z_2$, 
is the density representing the sensor deployment (recall that  
nodes are assumed to be uniformly distributed in the region
hence their density is constant and equal to 1).
Using~(\ref{eq:SNR}), we obtain:
\begin{eqnarray} 
f_{x}(z_1,z_2) 
&=& \frac{\exp\left(-a(z_1^2+z_2^2)\right)}{\iint_{\Hc}\exp\left(- a(z_1^2+z_2^2)\right) \dd z_1 \dd z_2} \non
&=& b\exp\left(-a(z_1^2+z_2^2)\right)
\label{eq:f_xi}
\end{eqnarray}
where
\[ b^{-1} = \iint_{\Hc}\exp\left(- a(z_1^2+z_2^2)\right) \dd z_1 \dd z_2 = \frac{\pi}{a}\erf^2\left(\sqrt{\frac{a}{4}}\right) \]

In order to compute~(\ref{eq:eta}), we need the function $g_{x}(y)$,
i.e., the density of $f_{x}(z_1,z_2)$. Note that $f_{x}(z_1,z_2)$
is circularly symmetric with respect to $(z_1,z_2)=(0,0)$.  Let $y$ be
the value of density of the sampling points at distance 
$d_y=\sqrt{z_1^2+z_2^2}$ from
the sink.  Then, from~(\ref{eq:f_xi}) we obtain
$d_y=\sqrt{\frac{1}{a}\log\frac{b}{y}}$, thus the network area where
the density is lower than $y$ is given by
\[ G_{x}(y) = 1-\pi d_y^2 \]
for $0\le d_y \le 1/2$, i.e.,  $b\ee^{-a/4} \le y<b$.
For $1/2<d_y<\sqrt{2}/2$, it is possible to show that
\[  G_{x}(y) = 1-\sqrt{4d_y^2-1}-d_y^2\left(\pi-4\cos^{-1}\frac{1}{2d_y} \right) \] 
In conclusion,
\begin{equation}
G_{x}(y) = \left\{
\begin{array}{ll}
1           &  y/b \ge 1 \\
1-\pi d_y^2 &  \ee^{-a/4} \le y/b < 1 \\
1-\sqrt{4d_y^2-1}-d_y^2\left(\pi-4\cos^{-1}\frac{1}{2d_y} \right) &  \ee^{-a/2} \le y/b < \ee^{-a/4} \\
0   & y/b<  \ee^{-a/2} 
\end{array}
\right.
\end{equation}
and
\begin{equation}
g_{x}(y) = \left\{
\begin{array}{ll}
0           &  y/b \ge 1 \\
\frac{\pi}{ay} &  \ee^{-a/4} \le y/b < 1 \\
\frac{1}{ay}\left(\pi-4\cos^{-1}\frac{1}{2d_y} \right) &  \ee^{-a/2} \le y/b < \ee^{-a/4} \\
0   & y/b <  \ee^{-a/2} 
\end{array}
\right.
\label{eq:g_xi}
\end{equation}
Since $|\Ac|=1$, then the asymptotic MSE can be obtained by computing
\begin{eqnarray}
\MSEinf=\eta_x(2,\beta,\gamma/\beta) = \int_{\ee^{-a/2}}^b g_x(y) \eta_u\left(2, \frac{\beta}{y}, \frac{\gamma y}{\beta}\right)\dd y
\label{eq:eta_xi_2}
\end{eqnarray}

\insertfig{0.8}{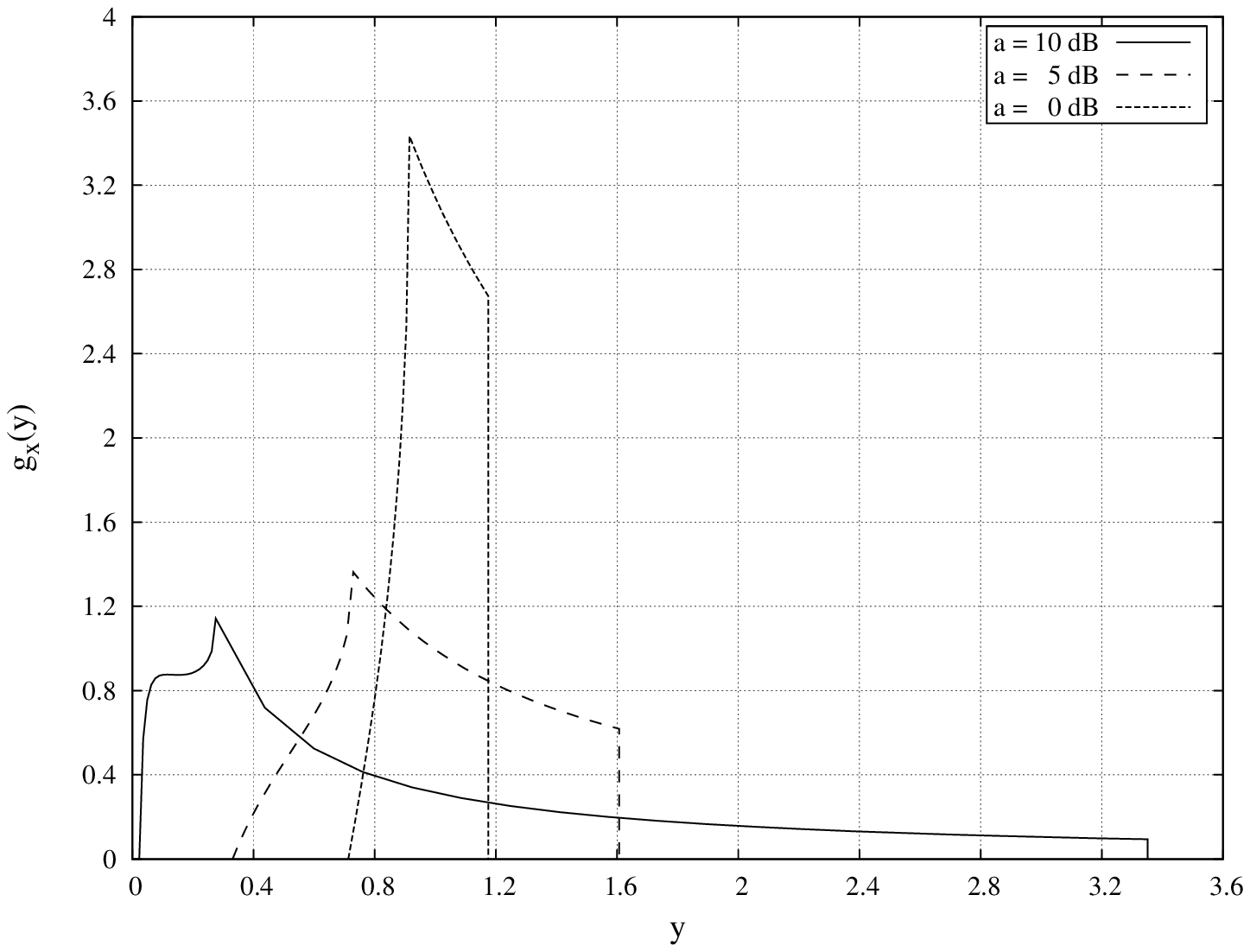}{Transmissions in presence of fading: density 
$g_x(y)$ for $a=0,5,10$~dB, i.e., for different values of the 
SNR threshold $\tau$.}{fig:g}

Figure~\ref{fig:g} shows the density $g_x(y)$ for $a=0,5,10$~dB. 
Note that $a=\tau/s$, thus for a fixed $s$ (i.e.,
the signal to noise ratio at distance $D$ in the absence of fading)
the parameter $a$ is proportional to the SNR threshold $\tau$. In particular,
as $\tau$ decreases, the probability that a message successfully
reaches the destination increases and, thus, the spatial
distribution of correctly received samples, $f_x(z_1,z_2)$, tends to
the uniform distribution $f_u(z_1,z_2)$. As a consequence, the density of
$f_{x}(z_1,z_2)$, i.e., $g_x(y)$, for $a=0$ and 5~dB 
is concentrated close to $y=1$. 
However, for high values of $\tau$, messages originated from
sensor nodes located far from the sink are successfully received
with low probability. Thus, $g_x(y)$ shows a significant 
probability mass around $y=0$.

\insertfig{0.8}{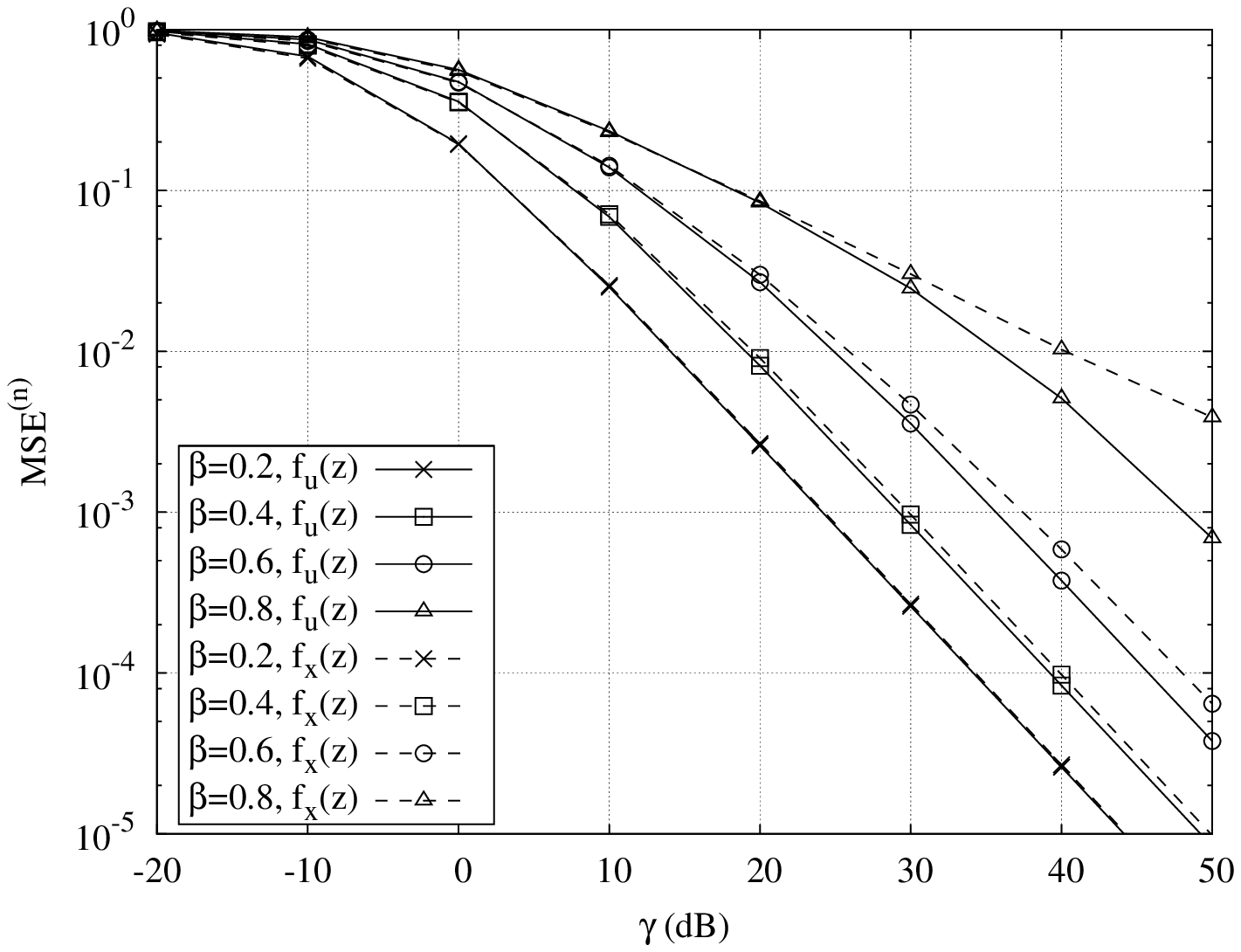}{MSE of the reconstructed field in absence
($f_u(\zv)$) and in presence ($f_x(\zv)$) of fading, as the signal to noise 
ratio on the sensor measurements varies.}{fig:mse}

Figure~\ref{fig:mse} shows the effect of the fading channel on the MSE
of the reconstructed field (dashed lines), and compares the obtained
results with the MSE obtained in absence of fading (solid lines).  The
plot considers different values of $\beta$, namely,
$\beta=0.2,0.4,0.6,0.8$, and $a=5$~dB. The MSE is plotted versus the
signal to noise ratio on the sensor measurements, $\gamma$.  The
curves have been obtained by numerically
computing~(\ref{eq:eta_xi_2}), where $g_x(y)$ is given
by~(\ref{eq:g_xi}) and $\eta_u(2,\beta,\gamma/\beta)$ is replaced by
$\eta^{(n)}_u(2,\beta_{n,m},\gamma/\beta_{n,m})$, with $n=10$.  Recall that the
analytic expression of the LSD $f_{\lambda,u}(d,\beta,z)$ is unknown,
hence in the numerical results we considered the AESD
$f^{(n)}_{\lambda,u}(d,\beta_{n,m},z)$ instead. We observe that for
low values $\beta$, in spite of the presence of fading, the sink node
still receives a large number of samples from the sensors, hence the
degradation of the MSE shown in Figure~\ref{fig:mse} is negligible. On the
contrary, for $\beta> 0.4$ (i.e., for a larger value of the ratio of
the number of harmonics composing the approximated field to the number 
of sensors), the reconstruction performance degrades
significantly and this is particularly evident in presence of high values
of $\gamma$.

\insertfig{0.8}{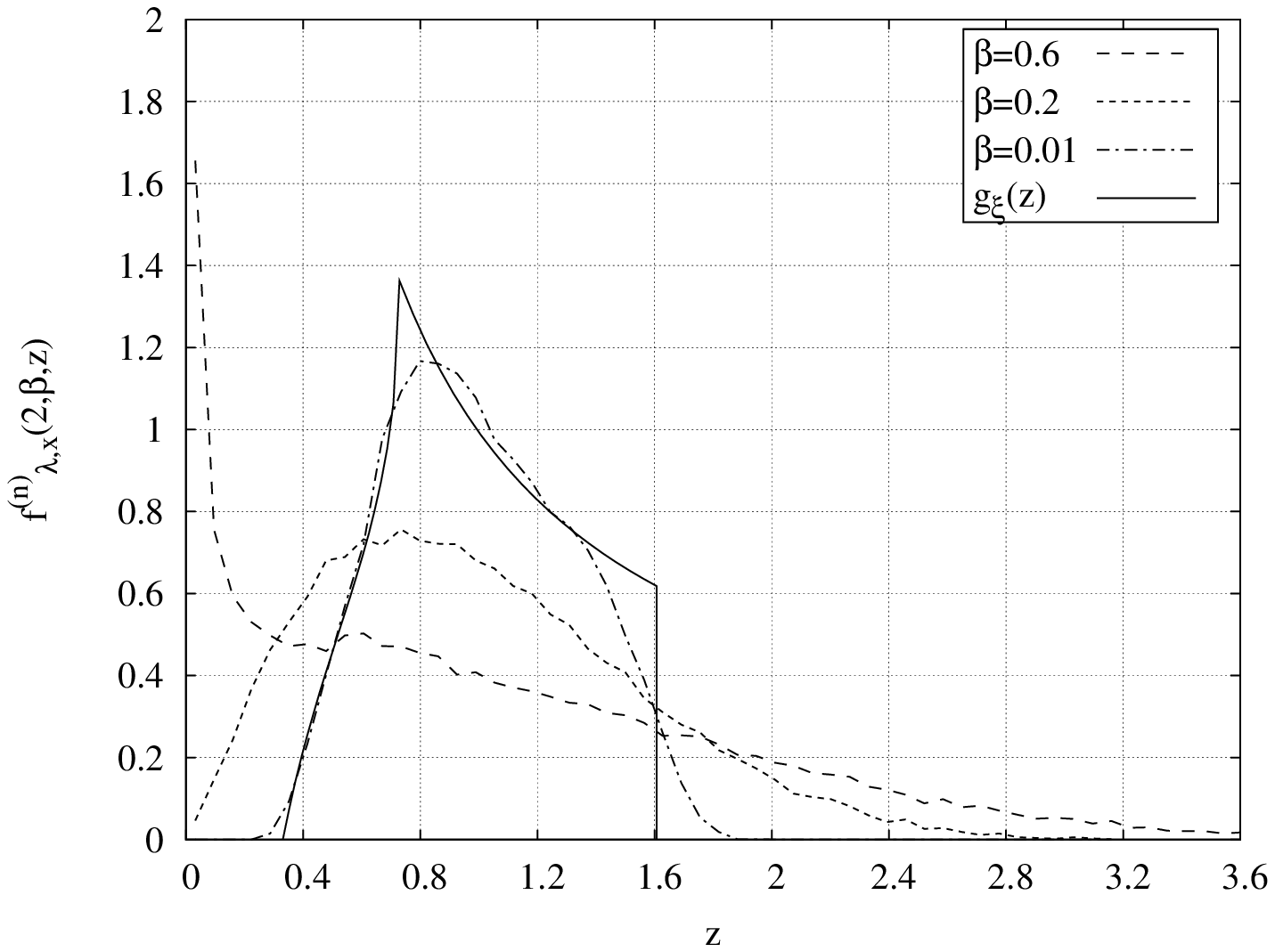}{Massively dense networks: 
empirical function $f^{(n)}_{\lambda,x}(2,\beta,z)$ 
in presence of fading, 
with $a=5$~dB and $|\Ac|=1$. 
The curves obtained for different values of $\beta$ are compared
with the density $g_{x}(z)$.}{fig:dense}

In the case of massively dense networks, the LSD
of $\Vm\Vm\Herm$ is given by (\ref{eq:cor1}) and from~(\ref{eq:cor3.2}) 
we know that
the MSE tends to 0 as $\beta \rightarrow \infty$. This result is confirmed by
the plot in Figure~\ref{fig:dense}, which shows the AESD $f^{(n)}_{\lambda,x}(2,\beta_{n,m},z)$, for
$|\Ac|=1$, $a=5$~dB, and $n=10$. The behavior of such a function is
compared with the density $g_{x}(z)$ as $\beta$ varies. We note
that, as $\beta$ decreases, the matching between
$f^{(n)}_{\lambda,x}(2,\beta_{n,m},z)$ and $g_{x}(z)$ improves,
and the latter represents an excellent approximation already for
$\beta=0.01$, as predicted by the result in~(\ref{cor1}).


\subsection{Measurements gathering through contention-based channel access}

In environmental monitoring applications, it is often desirable to vary
the resolution level with which the field measurements are taken
over the region under observation, depending on the field variations and 
the interest level of the different locations~\cite{Perillo04,Willett04}. 
It follows that the number of samples generated by the sensors network 
(i.e., the offered traffic load)  varies in the spatial 
domain. 
 
To represent such a scenario,  we consider a wireless sensor network 
whose nodes are uniformly deployed over a square region. 
We also identify $L$ areas, $A_i$ $i=1,\ldots,L$,
each corresponding to a different value of the offered traffic load.
As often assumed in the literature (see e.g.,~\cite{cluster,clusterH})
and widely applied in the practice, 
the network is divided into clusters
and a hierarchy of clusters is created. 
More specifically, at the first hierarchical layer, layer 1, the sensors 
are grouped into clusters, each of which is 
controlled by a cluster-head. 
The cluster-head is in charge of handling all traffic 
packets it receives from the nodes.
At a given layer $h>1$ of the hierarchy, 
the cluster-heads are grouped into clusters on their turn and 
forward the traffic to their parent cluster-head.
At the highest layer, layer $H$, we have only one cluster whose 
cluster-head coincides with the sink node.
Without loss of generality, we assume that the cluster at the $H$-th layer 
is composed of $L$ cluster-heads, each handling the traffic 
generated within one of the $L$ areas defined above.

As for the medium access control (MAC) layer, we consider that the nodes
implement the IEEE 802.15.4 standard 
specifications for wireless sensor networks~\cite{802.15.4}. 
In particular, all nodes  within a cluster are in radio visibility 
of each other and use the  slotted
carrier-sense multiple-access/collision avoidance  
 (CSMA/CA) technique~\cite{802.15.4}. This is a contention-based scheme 
and transmissions may fail if two or more sensors
access the channel at the same time. Inter-cluster interference 
is instead avoided by assigning  different frequency 
channels to neighboring clusters. 
We consider that packets, whose transmission fails, are discarded.

In order to derive the probability that a packet transmission fails within
a cluster due
to collision, we use the Markov chain  model presented in~\cite{collision}.
We denote by $m_{i,h}$ the average number of sensors belonging to the generic
cluster at the $h$-th layer of the hierarchical architecture, in area $A_i$
($i=1,\ldots,L$ and $h=1,\ldots,H$). 
Similarly, we define $\lambda_{i,h}$ as the average traffic load per node, 
again within the generic cluster at the $h$-th layer, in area $A_i$. 
Then, we set the size of the packet payload to 32~bytes, 
and  the value of the other parameters as in~\cite{collision}. 
Under this setting, 
we compute the value of the collision probability within
the generic cluster at layer $h$, in area $A_i$, 
as a function of $m_{i,h}$ and  $\lambda_{i,h}$, i.e., 
$P_c(i,h)$~\cite{collision}. 
Furthermore, we observe that at the generic layer $h$, with $1<h\leq H$, 
a node, which acts as cluster-head  at layer $h-1$ in area $A_i$, will
have a traffic load equal to 
$\lambda_{i,h}=m_{i,h-1}\lambda_{i,h-1} \left [1-P_c(i,h-1)\right ]$.

It follows that the probability that a packet 
is successfully delivered to the corresponding $h$-layer cluster-head 
within area $A_i$ ($i=1,\ldots,L$) can be obtained as 
$P_s(i,h)=1-P_c(i,h)$.
Then, the probability that a measurement generated
by a sensor located in $A_i$ ($i=1,\ldots,L$) is successfully 
delivered to the sink is given by: 
\[
P_s(i)=1-\prod_{h=1}^H P_s(i,h)  .
\]

Next, denoting by $|A_i|$ the measure of $A_i$, we define 
\[ p_s(i) = \frac{P_s(i)}{\sum_{i=1}^L |A_i| P_s(i)} \]
as the normalized probability that a message is successfully
delivered to the sink. Then, the spatial density of the sensors successfully
sending their message is as follows:
\[  
f_{x}(z_1,z_2) = p_s(i) \quad \quad \forall (z_1,z_2)\in A_i,~i=1,\ldots,L .
\]
The density of $f_{x}(z_1,z_2)$ is therefore given by
\[ g_{x}(y) = \sum_{i=1}^L |A_i|\delta\left(y-p_s(i)\right) .\]
and the asymptotic MSE is given by
\[ \MSEinf = \eta_x(2,\beta,\frac{\gamma}{\beta}) = 
\sum_{i=1}^L |A_i|\eta_u\left(2,\frac{\beta}{p_s(i)},\frac{\gamma}{\beta} p_s(i) \right) \]

\insertfig{0.8}{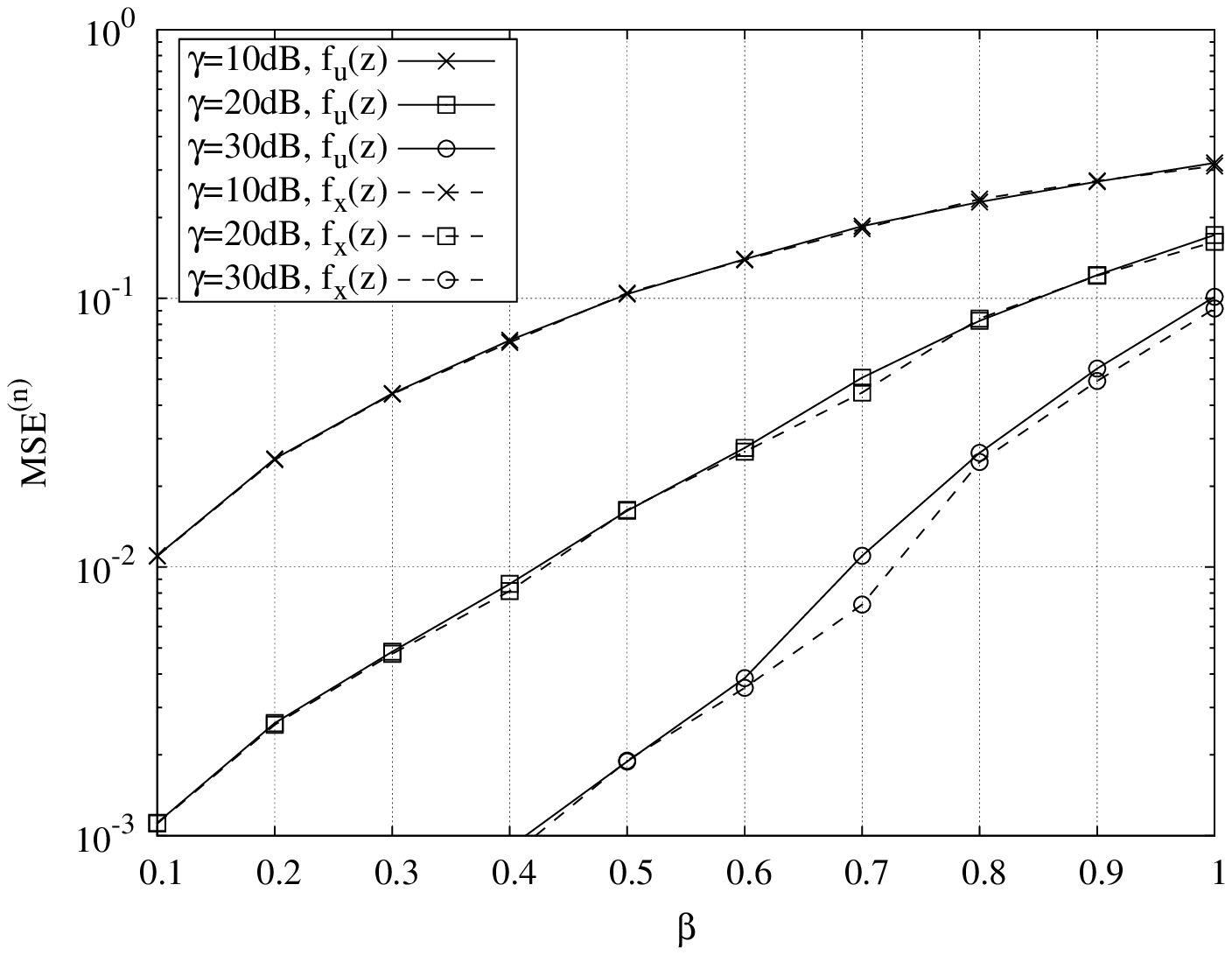}{Comparison between the case where transmission
collisions are taken into account ($f_x(\zv)$) and the case where all measurements successfully
reach the sink ($f_u(\zv)$). The MSE is shown as a function of $\beta$ and for different values of signal-to-noise ratio ($H=3$, $|A_i|=1/4, \forall i$, 
$\lambda_{1,1} =10^{-3}$, $\lambda_{2,1}=2\cdot 10^{-4}$, $\lambda_{3,1} =2\cdot 10^{-4}$, $\lambda_{4,1}=2\cdot 10^{-5}$).}{fig:lambda2}

\insertfig{0.8}{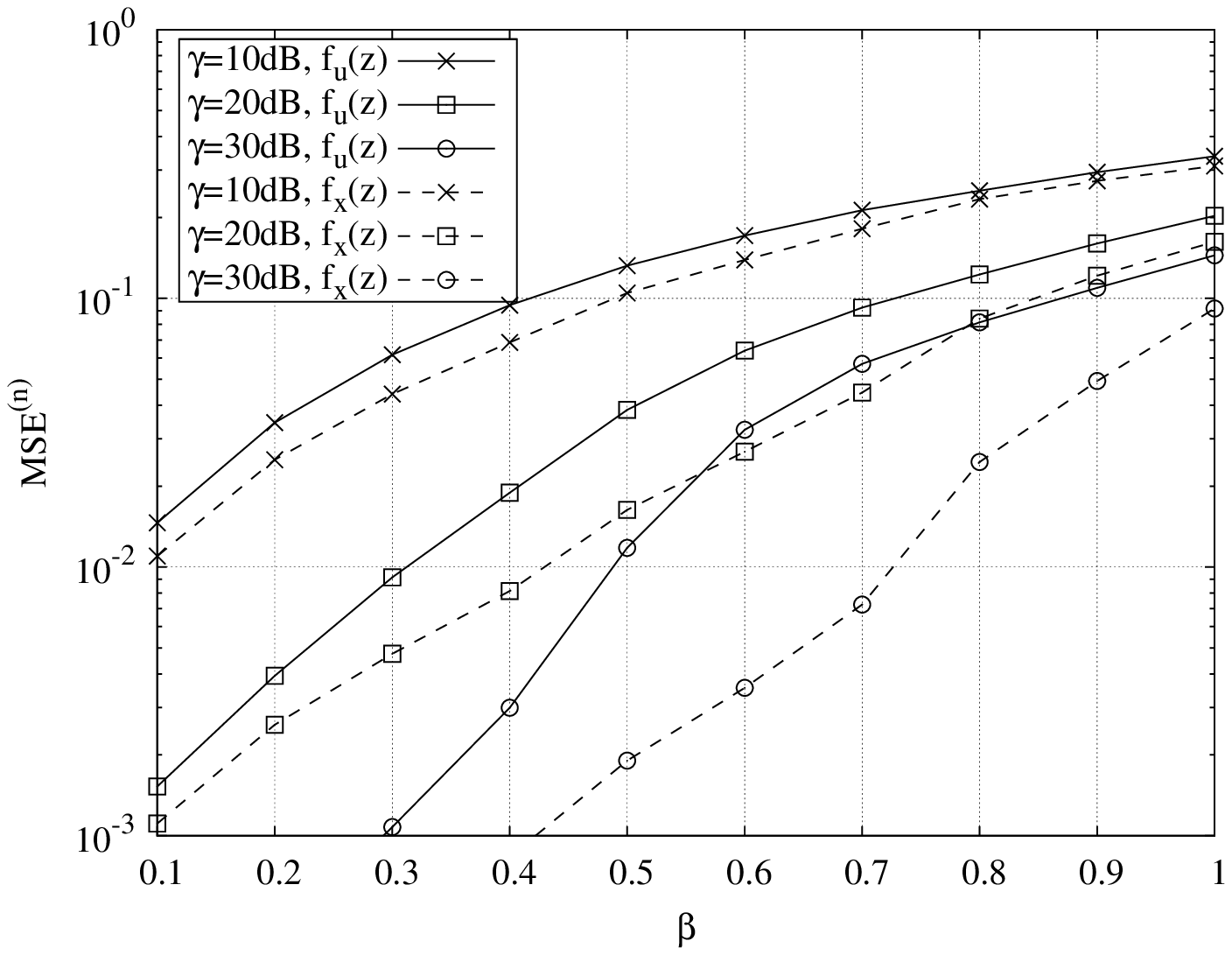}{Comparison between the case where transmission
collisions are taken into account ($f_x(\zv)$) and the case where all measurement transmissions are successful ($f_u(\zv)$). The MSE is shown as $\beta$ varies and for different values of signal-to-noise ratio ($H=3$, $|A_i|=1/4, \forall i$, 
$\lambda_{1,1} =5\cdot 10^{-3}$, $\lambda_{2,1}=10^{-3}$, $\lambda_{3,1} =10^{-3}$, $\lambda_{4,1}=10^{-4}$).}{fig:lambda1}

Figures~\ref{fig:lambda2} and \ref{fig:lambda1} show the impact of 
collisions due to the contention-based channel access, on the
quality of the reconstructed field. In particular, they compare the MSE of the
reconstructed field when collisions are taken into account ($f_x(\zv)$) with
the one obtained in the idealistic case where all messages (measurements) 
sent by the sensors successfully reach the sink ($f_u(\zv)$).
The results refer to a square region of unitary side, where there are four
areas of equal size ($|A_i|=1/4$, $i=1,\ldots,4$) but corresponding 
to different resolution levels in the measurements collection (i.e.,
they are characterized by different traffic loads);
the number of hierarchical levels is set to $H=3$.
We set $\lambda_{1,1} =10^{-3}$, $\lambda_{2,1}=2\cdot 10^{-4}$, $\lambda_{3,1} =2\cdot 10^{-4}$, $\lambda_{4,1}=2\cdot 10^{-5}$ in Figure~\ref{fig:lambda2}, and
a higher traffic load in Figure~\ref{fig:lambda1}, i.e., $\lambda_{1,1} =5\cdot 10^{-3}$, $\lambda_{2,1}=10^{-3}$, $\lambda_{3,1} =10^{-3}$, $\lambda_{4,1}=10^{-4}$. 

Looking at the plots, we observe that both $\beta$ and $\gamma$ have a significant impact of the obtained MSE, with the MSE increasing as $\beta$ grows and smaller values of $\gamma$ are considered. 
Most interestingly, by comparing the two figures,
we can see that as the traffic load, hence the collision probability, increases, the performance derived taking into account the contention-based channel access significantly differs from
the idealistic one. Furthermore, the latter effect is particularly evident as
$\gamma$ increases, since the higher the signal-to-noise ratio, the more
valuable the samples sent by the sensors toward the sink.

\section{Conclusion\label{sec:conclusions}}
We studied the performance of a wireless network whose nodes
sense a multi-dimensional field and transfer their measurements to
a sink node. As often happens in practical cases, we assumed the sensors
to be randomly deployed over (the whole or only a portion of) the 
region of interest, and that their measurements may be lost due to
fading or transmission collisions over the wireless channel.
We modeled the sampling system through
a multi-folded Vandermonde matrix $\Vm$ and, by using asymptotic analysis, 
we approximated the MSE of the field, which the sink node reconstructs
from the received sensor measurements with the $\eta$-transform of 
$\Vm\Vm\Herm$. 

Our results clearly indicate that the percentage of region where sensors cannot
be deployed must be extremely small if an accurate field estimation has to be obtained. Also, the effect of losses due to fading or transmission collisions
can be greatly mitigated provided that a suitable value for the
ratio between the number of harmonics approximating the field
bandwidth and the number of sensors is selected.

\appendices

\section{Proof of Theorem~\ref{th:1}}
\label{app:th1}
The $p$-th moment of the asymptotic eigenvalue distribution of
$\Vm\Vm\Herm$ can be expressed as~\cite{TSP2}
\[ M_{p,d,\beta,x}= \lim_{n,m \rightarrow \infty}\EE\left[\tracen\left\{\left(\Vm\Vm\Herm\right)^p\right\}\right] \]
where $\tracen \{\cdot\}$ is the normalized matrix trace operator.
The matrix power can be expanded as a multiple sum over the entries of
$\Vm$:
\[  M_{p,d,\beta,x}= \lim_{n,m \rightarrow \infty}\frac{1}{m^p n^{d}}\sum_{\ellv_1,\ldots,\ellv_p}\sum_{q_1,\ldots,q_p}\EE\left[ \ee^{\jj 2\pi \ellv_1\Tran (\xv_{q_p}-\xv_{q_1}) }\cdots \ee^{-\jj 2\pi \ellv_p\Tran (\xv_{q_{p-1}}-\xv_{q_{p}})} \right]  \]
where $q_1,\cdots,q_p$, $q_i = 1,\ldots,m$ are integer indices and
$\ellv_1,\ldots,\ellv_p$,
$\ellv_i=[\ell_{i,1},\ldots,\ellv_{i,d}]\Tran$,
$\ellv_{i,j}=0,\ldots,n-1$ are the indices identifying the rows of $\Vm$.
Since, 
\[ \sum_{\ellv_i} \ee^{\jj 2 \pi \ellv\Tran \xv} = \sum_{\ell_{i,1},\ldots,\ell_{i,d}=0}^{n-1} \ee^{\jj 2 \pi \left(\ell_{i,1} x_1+ \cdots + \ell_{i,d} x_d\right)} = \prod_{j=1}^d\frac{1-\ee^{\jj 2 \pi n x_j}}{1-\ee^{\jj 2 \pi x_j}} \]
for $i=1,\ldots,p$ and the elements of $\xv$ are i.i.d., we have that
\[ M_{p,d,\beta,x}
= \lim_{n,m \rightarrow \infty}\frac{1}{m^p n^d} \sum_{q_1,\ldots,q_p} \EE\left[\prod_{i=1}^p\prod_{j=1}^d\frac{1-\ee^{\jj 2 \pi n \left(x_{q_i,j}-x_{q_{i+1},j}\right)}}{1-\ee^{\jj 2 \pi\left(x_{q_i,j}-x_{q_{i+1},j}\right)}}  \right] \non
\]
where the index $i$ is to be considered modulo $p$, i.e., $p+1 \equiv
1$.  As for the sum over the indices $q_1,\ldots,q_p$ we note that any
choice of $\qv=[q_1,\ldots,q_p]\Tran$ induces a partition $\omegav$ of
the set, $\Pc=\{1,\ldots,p\}$ in $k$ subsets $\Pc_1,\ldots,\Pc_k$,
$1\le k \le p$, under the equality relation~\cite{TSP2}.  In the
following, we denote by $\Omega_{p,k}$ the set of partitions of $\Pc$
in $k$ subsets, $1\le k \le p$. Since there are $m^k$ possible vectors
$\qv$ inducing a given partition $\omegav \in \Omega_{p,k}$, we can
write the $p$-th moment as
\begin{eqnarray}
M_{p,d,\beta,x}
&=&  \lim_{n,m \rightarrow \infty}\sum_{k=1}^p\sum_{\omegav \in \Omega_{p,k}} 
     \frac{\beta_{n,m}^{p-k}}{n^{d(p-k+1)}}\EE\left[\prod_{i=1}^p\prod_{j=1}^d
     \frac{1-\ee^{\jj 2 \pi n \left(x_{\omega_i,j}-x_{\omega_{i+1},j}\right)}}{1-\ee^{\jj 2 \pi\left(x_{\omega_i,j}-x_{\omega_{i+1},j}\right)}} \right] \non
&=&  \lim_{n,m \rightarrow \infty}\sum_{k=1}^p\beta_{n,m}^{p-k}\sum_{\omegav \in \Omega_{p,k}} \frac{\EE\left[\Phi_{\omegav} (\xv_1,\ldots,\xv_k) \right]}{n^{d(p-k+1)}}
\label{eq:moment2}
\end{eqnarray}
where $\Phi_{\omegav}(\xv_1,\ldots,\xv_k)
=\prod_{j=1}^dF_{\omegav}(x_{1j},\ldots,x_{kj})$,
$F_{\omegav}(x_{1j},\ldots,x_{kj})= \prod_{i=1}^p\frac{1-\ee^{\jj 2
    \pi n \left(x_{\omega_i,j}-x_{\omega_{i+1},j}\right)}}{1-\ee^{\jj
    2 \pi\left(x_{\omega_i,j}-x_{\omega_{i+1},j}\right)}}$, and
$\omega_i \in \{1, \ldots, k\}$ is the index of the subset of $\Pc$
containing $i$. Recall that $p+1\equiv 1$ and that
$\beta_{n,m}=n^d/m$.  Moreover, since the vectors $\xv$ are i.i.d., we
removed the dependence on the subscript $q$.

Following the same steps as in~\cite[Appendix H]{RyanDebbah}, we compute the limit  
\begin{eqnarray}
\lim_{n \rightarrow \infty}\frac{\EE\left[\Phi_{\omegav} (\xv_1,\ldots,\xv_k) \right]}{n^{d(p-k+1)}} 
&=&\lim_{n \rightarrow \infty}\int_{\Hc^k}f_x(\xv_1)\cdots f_x(\xv_k)\frac{\Phi_{\omegav}(\xv_1,\ldots,\xv_k)}{n^{d(p+1-k)}} \dd \xv_1 \cdots \dd \xv_k \non
&=&\lim_{n \rightarrow \infty}\int_{\Hc^k}f_x(\xv_1)\cdots f_x(\xv_k)\prod_{j=1}^d \frac{F_{\omegav}(x_{1j},\ldots,x_{kj})}{n^{p+1-k}} \dd x_{1j}\cdots \dd x_{kj} \nonumber
\end{eqnarray}
We then define $\xv_h  = [x_{h1}, \yv_h]$ where $\yv_h = [x_{h2}, \ldots,x_{hd}]$, for $h=1,\ldots,k$ 
and we integrate first with respect to the variables $x_{11},\ldots, x_{k1}$ obtaining
\begin{eqnarray}
\lim_{n \rightarrow \infty}\frac{\EE\left[\Phi_{\omegav} (\xv_1,\ldots,\xv_k) \right]}{n^{d(p-k+1)}} 
&=& \lim_{n \rightarrow \infty}\int_{[-\frac{1}{2},\frac{1}{2}]^{(d-1)k}} G_{\omegav}(\yv_1,\ldots,\yv_k)\prod_{j=2}^d \frac{F_{\omegav}(x_{1j},\ldots,x_{kj})}{n^{p-k+1}} \dd x_{1j}\cdots \dd x_{kj} \nonumber\end{eqnarray}
where
\begin{equation}
 G_{\omegav}(\yv_1,\ldots,\yv_k) = \int_{[-\frac{1}{2},\frac{1}{2}]^{k}}\frac{F_{\omegav}(x_{11},\ldots,x_{k1})}{n^{p-k+1}}f_x([x_{11}, \yv_1])\cdots f_x([x_{k1}, \yv_k])\dd x_{11}\cdots \dd x_{k1} 
\label{eq:G_omega}
\end{equation}
In~\cite[Appendix H]{RyanDebbah} it was shown that, because of the properties of $F_{\omegav}(x_{11},\ldots,x_{k1})$,
\[ \lim_{n \rightarrow \infty} \int_{\Bc_\epsilon}\frac{F_{\omegav}(x_{11},\ldots,x_{k1})}{n^{p-k+1}}\dd x_{11}\cdots \dd x_{k1} = 0 \]
where
\[\Bc_\epsilon=\left\{(x_{11},\ldots,x_{k1}) | |x_{h1}-x_{\ell 1}|>\epsilon, \mbox{for some} \,\,h,\ell \right\}\] 
for any $\epsilon>0$.  This means that the integral
in~(\ref{eq:G_omega}) can be limited to the $x_{11},\ldots,x_{k1}$ on
the diagonal where $x_{11}= \cdots =x_{k1}$. Therefore
\begin{eqnarray}
 \lim_{n \rightarrow \infty} G_{\omegav}(\yv_1,\ldots,\yv_k) &=& 
 \lim_{n \rightarrow \infty}\int_{[-\frac{1}{2},\frac{1}{2}]^{k}}\frac{F_{\omegav}(x_{11},\ldots,x_{k1})}{n^{p-k+1}} f_x([x_{k1}, \yv_1])\cdots f_x([x_{k1}, \yv_k])\dd x_{11}\cdots \dd x_{k1} \non
&=& \int_{[-\frac{1}{2},\frac{1}{2}]} \prod_{h=1}^k f_x([x_{k1}, \yv_h]) \lim_{n \rightarrow \infty}\left(\int_{[-\frac{1}{2},\frac{1}{2}]^{k-1}}\frac{F_{\omegav}(x_{11},\ldots,x_{k1})}{n^{p-k+1}}\prod_{h=1}^{k-1}\dd x_{h1}\right)  \dd x_{k1} \non
&=& v(\omegav) \int_{[-\frac{1}{2},\frac{1}{2}]} \prod_{h=1}^k f_x([x_{k1}, \yv_h]) \dd x_{k1} 
\end{eqnarray}
Note that the limit
\[v(\omegav) = \lim_{n \rightarrow \infty}\int_{[-\frac{1}{2},\frac{1}{2}]^{k-1}}\frac{F_{\omegav}(x_{11},\ldots,x_{k1})}{n^{p-k+1}}\prod_{h=1}^{k-1}\dd x_{h1}\] 
does not depend on $x_{k1}$ and the coefficient $v(\omegav) \in [0,1]$ is described in~\cite{TSP2}.

Next, iterating this procedure by integrating over the variables, $x_{1j},\ldots, x_{kj}$, $j=2,\ldots,d$ we finally get
\begin{eqnarray} \lim_{n \rightarrow \infty}\frac{\EE\left[\Phi_{\omegav} (\xv_1,\ldots,\xv_k) \right]}{n^{d(p-k+1)}} 
&=& v(\omegav)^d \int_{[-\frac{1}{2},\frac{1}{2}]^d} f_x(x_{k1}, \ldots ,x_{kd})^k \dd x_{k1} \cdots \dd x_{kd} \non
&=& v(\omegav)^d \int_{\Hc} f_x(\xv_k)^k \dd \xv_k \non
&=& v(\omegav)^d I_k
\end{eqnarray}
where we defined $I_k=\int_{\Hc}f_x(\xv_k)^k\dd \xv_k$.  It follows that
\[ M_{p,d,\beta,x} = \sum_{k=1}^p\beta^{p-k} I_k \sum_{\omegav \in \Omega_{p,k}}v(\omegav)^d \]
which proves the theorem. Note that when the entries of
$\xv_q=[x_{q1},\ldots,x_{qd}]\Tran$ are independent with continuous
distribution $f_{x,j}(z_j)$ such that $f_x(\xv) = \prod_{j=1}^d
f_{x_j}(x_j)$, we have $I_k = \prod_{j=1}^d I_{k,j}$ with
$I_{k,j}=\int_{[-1/2,1/2]} f_{x_j}(x)^k\dd x$.

\section{Proof of Theorem~\ref{th:2}}
\label{app:th2}
From Theorem~\ref{th:1} and the definition of $I_k$, we have that
\begin{eqnarray} 
M_{p,d,\beta,x} &=& \sum_{k=1}^p\beta^{p-k} I_k \sum_{\omegav \in \Omega_{p,k}}v(\omegav)^d \non
&=&\int_{\Hc}\sum_{k=1}^p\beta^{p-k} f_x(\zv)^k \sum_{\omegav \in \Omega_{p,k}}v(\omegav)^d \dd \zv  .
\label{eq:moment_psi}
\end{eqnarray}
Next, we define the set $\Ac$ where $f_x(\zv)$ is strictly positive
as
\[ \Ac = \left\{ \zv \in \Hc | f_x(\zv)>0  \right\} \]
Note that for $\zv\in \Hc\setminus\Ac$ the contribution to the
  integral in~(\ref{eq:moment_psi}) is zero. Thus,
\begin{eqnarray} 
M_{p,d,\beta,x} &=& 
\int_{\Ac}\sum_{k=1}^p\beta^{p-k} f_x(\zv)^k \sum_{\omegav \in \Omega_{p,k}}v(\omegav)^d \dd \zv \non
&=&\int_{\Ac}f_x(\zv)^p \sum_{k=1}^p\beta^{p-k} f_x(\zv)^{k-p} \sum_{\omegav \in \Omega_{p,k}}v(\omegav)^d \dd \zv \non
&=&\int_{\Ac}f_x(\zv)^p  \sum_{k=1}^p\beta'(\zv)^{p-k} \sum_{\omegav \in \Omega_{p,k}}v(\omegav)^d \dd \zv \non
&=&\int_{\Ac}f_x(\zv)^p M_{p,d,\beta'(\zv),u} \dd \zv 
\label{eq:moments_psi2}
\end{eqnarray}
where for any $\zv \in \Ac$, $M_{p,d,\beta'(\zv)}$ is the $p$-th moment
of $\Vm\Vm\Herm$ when the phases are uniformly
distributed in $\Hc$ and the ratio $\beta'(\zv)$ is given by
\[ \beta'(\zv) = \frac{\beta}{f_x(\zv)} \]
Note also that~(\ref{eq:moments_psi2}) holds for $p\ge 1$ since, by
definition, the zero-th moment of any distribution is equal to
1. Expression~(\ref{eq:moments_psi2}) allows us to write the moments of
$\Vm\Vm\Herm$ for any distribution $f_x(\zv)$, given the moments for
uniformly distributed phases.  Likewise, it is possible to describe
the LSD of $\Vm\Vm\Herm$, for any continuous $f_x(\zv)$, in terms of
the LSD obtained for uniformly distributed phases.  Indeed, let us
denote the Laplace transform of $f_{\lambda,x}(d,\beta,z)$ by
$L_{\lambda,x}(d,\beta,s)$ if it exists. Then, whenever the sum
converges
\[  L_{\lambda,x}(d,\beta,s) = \sum_{p=0}^{\infty} \frac{s^p}{p!}M_{p,d,\beta,x} \]
Since $M_{0,d,\beta,x}=1$ for any distribution,
using~(\ref{eq:moments_psi2}) we obtain:
\begin{eqnarray} 
 L_{\lambda,x}(d,\beta,s)
&=& 1+\sum_{p=1}^{\infty}\frac{s^p}{p!}\int_{\Ac}f_x(\zv)^p M_{p,d,\beta'(\zv),u} \dd \zv \non
&=& 1-|\Ac|+|\Ac|+\int_{\Ac}\sum_{p=1}^{\infty}\frac{f_x(\zv)^p s^p}{n!}M_{p,d,\beta'(\zv),u}\dd \zv \non
&=& 1-|\Ac|+\int_{\Ac}\sum_{p=0}^{\infty}\frac{f_x(\zv)^p s^n}{n!}M_{p,d,\beta'(\zv),u}\dd \zv \non
 &=& 1-|\Ac|+\int_{\Ac}L_{\lambda,u}\left(d,\beta'(\zv),f_x(\zv)s\right)\dd \zv \non
\end{eqnarray}
where $|\Ac|$ is the measure of $\Ac$ and $L_{\lambda,u}(d,\beta,s)$
is the Laplace transform of $f_{\lambda,u}(d,\beta,z)$. By using the
properties of the Laplace transform and by taking its inverse, we
finally get
\begin{eqnarray} 
f_{\lambda,x}(d,\beta,z) = \left(1-|\Ac|\right)\delta(z)+\int_{\Ac}\frac{1}{f_x(\zv)}
f_{\lambda,u}\left(d,\frac{\beta}{f_x(\zv)},\frac{z}{f_x(\zv)}\right)\dd \zv  .
\label{eq:distribution_psi}
\end{eqnarray}

We can rewrite the second term of~(\ref{eq:distribution_psi}) by
defining the cumulative density function 
\[ G_x(y) = \frac{1}{|\Ac|}\left|\left\{\zv \in \Ac | f_x(\zv)\le y \right\}\right| \] 
for $y >0$. By using the corresponding probability
  density function, $g_x(y)$, and Lebesgue integration, we can
 rewrite~(\ref{eq:distribution_psi}) as in~(\ref{eq:th2}).

\section{Corollary~\ref{cor-scal}}
\label{app:cor-scal}
From the result in~(\ref{eq:th2}) and from the assumption
$f_{x}(\zv)>0$ $\forall\zv \in \Hc$ (i.e., $|\Ac|=1$), we have
\[ f_{\lambda,x}(d,\beta,z) =  \int_0^{\infty}\frac{g_x(y)}{y}f_{\lambda,u}\left(d,\frac{\beta}{y},\frac{z}{y}\right)\dd y \]
Then, from the definition of $f_{x'}(\zv)$ given
in~(\ref{eq:fxi'_scaled}) it follows that $G_{x'}(y) = G_x(cy)$
and, by consequence, $g_{x'}(y) = c g_x(cy)$. Therefore,
from~(\ref{eq:f-scal}) we have:
\begin{eqnarray}
f_{\lambda,x'}(d,\beta,z) 
&=& (1-c)\delta(z)+ c\int_0^{\infty}\frac{g_{x'}(y)}{y}f_{\lambda,u}\left(d,\frac{\beta}{y},\frac{z}{y}\right)\dd y \non
&=& (1-c)\delta(z)+ c^2\int_0^{\infty}\frac{g_{x}(cy)}{y}f_{\lambda,u}\left(d,\frac{\beta}{y},\frac{z}{y}\right)\dd y \non
&=& (1-c)\delta(z)+ c^2\int_0^{\infty}\frac{g_{x}(y)}{y}f_{\lambda,u}\left(d,\frac{c\beta}{y},\frac{cz}{y}\right)\dd y \non
&=& (1-c)\delta(z)+ c^2 f_{\lambda,x}(d,c\beta,cz)\nonumber
\end{eqnarray}

\section{Corollary~\ref{cor1}}
\label{app:cor1}
From the expression of the moments given in Theorem~\ref{th:1} and the
results in~\cite{TSP2}, it is easy to show that for uniformly
distributed phases, we have:
\[ \lim_{\beta\rightarrow 0} M_{p,d,\beta,x} = 1\]
for any $p \ge 0$. It immediately follows that 
\[ \lim_{\beta\rightarrow 0}f_{\lambda,u}(d,\beta,z) = \delta(z-1) \]
where $\delta(z)$ is the Dirac's delta function. By applying this
result to~(\ref{eq:th2}), we get
\begin{eqnarray}
\lim_{\beta\rightarrow 0}f_{\lambda,x}\left(d,\frac{\beta}{y},\frac{z}{y}\right) 
&=& \left(1-|\Ac|\right)\delta(z)+ |\Ac| \int_{0}^{\infty}\frac{g_x(y)}{y}\delta\left(\frac{z}{y}-1\right)\dd y \non
&=& \left(1-|\Ac|\right)\delta(z)+ |\Ac| \int_{0}^{\infty}\frac{g_x(z/w)}{w}\delta(w-1)\dd w \non
&=& \left(1-|\Ac|\right)\delta(z)+ |\Ac| g_x(z)  .
\label{eq:distribution_psi3}
\end{eqnarray}

\section{Corollary~\ref{cor2}}
\label{app:cor2}
By using the definition of the $\eta$-transform and the result
in~(\ref{eq:th2}), we  obtain:
\begin{eqnarray}
\eta_{x}(d,\beta,\gamma) 
&=& \EE\left[\left(\gamma\lambda+1\right)^{-1}\right] \non
&=& \int_0^{\infty} \frac{1}{\gamma z+1}f_{\lambda,x}(d,\beta,z) \dd z \non
&=& \int_0^{\infty} \frac{1-|\Ac|}{\gamma z+1}\delta(z) 
    + |\Ac| \int_0^{\infty}\frac{g_x(y)}{y} \int_0^{\infty }\frac{1}{\gamma z+1}f_{\lambda,u}\left(d,\frac{\beta}{y},\frac{z}{y}\right)\dd z \dd y \non
&=& 1-|\Ac|+ |\Ac| \int_0^{\infty} g_x(y) \int_0^{\infty }\frac{1}{\gamma y z+1}f_{\lambda,u}\left(d,\frac{\beta}{y}, z\right)\dd z \dd y \non
&=& 1-|\Ac|+ |\Ac| \int_0^{\infty} g_x(y) \eta_u\left(d,\frac{\beta}{y}, \gamma y\right) \dd y .
\end{eqnarray}
Then, by considering that $\MSEinf = \eta_{x}(d,\beta,\gamma/\beta)$, 
the expression of the asymptotic MSE 
immediately follows. 

\section{Corollary~\ref{cor3}}
\label{app:cor3}
In Appendix~\ref{cor1} we have shown that $\lim_{\beta\rightarrow
  0}f_{\lambda,u}(d,\beta,z) = \delta(z-1)$. From the definition of
the $\eta$-transform, it follows that $\lim_{\beta\rightarrow
  0}\eta_{u}(d,\beta,\gamma) = \frac{1}{\gamma +1}$ and
$\lim_{\beta\rightarrow 0}\eta_{u}(d,\beta,\gamma/\beta) =
0$. Thus, from~(\ref{eq:eta}) we have:
\begin{eqnarray} \lim_{\beta\rightarrow 0} \eta_x(d,\beta,\gamma) 
&=& 1-|\Ac| +|\Ac| \int_0^{\infty}g_x(y)\left( \lim_{\beta\rightarrow 0} \eta_{u}\left(d,\frac{\beta}{y},\gamma y\right)\right)\dd y \non
&=& 1-|\Ac| +|\Ac| \int_0^{\infty}\frac{g_x(y)}{\gamma y +1} \dd y \non
&=& 1-|\Ac| +|\Ac| \eta_g(\gamma)
\end{eqnarray}
where we defined $\eta_g(\gamma) = \int_0^{\infty}\frac{g_x(y)}{\gamma y +1} \dd y$.
As a consequence,
\begin{eqnarray} \lim_{\beta\rightarrow 0} \eta_x(d,\beta,\gamma/\beta) 
&=& 1-|\Ac| +|\Ac| \int_0^{\infty}g_x(y)\left( \lim_{\beta\rightarrow 0} \eta_{u}\left(d,\frac{\beta}{y},\frac{\gamma y}{\beta}\right)\right)\dd y \non
&=& 1-|\Ac| .
\end{eqnarray}

\end{document}